**What does the data of the Brexit referendum really say?**

**Nicholas Donaldson[1], Nora Donaldson[2] and Grace Yang[3]**

[1] Freelance video journalist (previously with Latest TV Brighton) and Research Assistant.

[2] Honorary Professor in Statistics at King's College Hospital NHS Trust, UK and Visiting Professor at Stony Brook University, USA. (Retired professor of Biostatistics. King's College London).

[3] Emeritus Professor in Statistics at the University of Maryland.

**Abstract**

The Brexit referendum took place in the UK in June, 2016. The unweighted percentage of leavers over the whole population was 51.9%. In this paper, first, we demonstrate that a 52%-48% split represents only a spurious difference, not sufficiently different from a 50-50 split to claim a majority for either side. Second, on this basis of the unweighted percentage, statement like *"The country voted to leave the EU…"* were made. When a statement about a population is made based on a subset of it (the turnout rate for Brexit was only 72% and therefore 37% of the eligible population voted Leave), it comes with an element of uncertainty that should not be ignored. The unweighted average disregards, not only between-region heterogeneity but also within-region variability. Our analysis, controlling for both, finds that the split of the Brexit is of negligible material significance and do not indicate majority for either side.

## Introduction

The Brexit referendum took place in the United Kingdom (UK) on the 23rd of June 2016. It was a single question asking respondents to choose one of two possible options, namely, whether the UK should remain or leave the European Union (EU). The result of the referendum was given in terms of the split 51.9%-48.1% (51.9% of the valid votes chose leave and 48.1% chose Remain), declaring a victory for Leave [1]. We address two fundamental issues in this paper.

The first issue is whether a 52%-48% split is sufficiently different from a 50%-50% split to substantiate the claim that there was a majority for one of the campaigns. Common sense indicates that a 60%-40% split (i.e. a 10% difference over 50%) would have been a definite conclusive result implying majority, and that a 50%-50% split would have been a perfectly even split implying, conclusively, no majority for either campaign. And perhaps common sense is all that is required to realize that a 52%-48% split represents a difference of no material or practical significance; that a 2% above 50% is not sufficiently large as to imply a majority for either campaign. This is precisely what was suggested in the public domain by two individuals in advance of the referendum, at the time when the polls were predicting such a split in favour of Remain. In fact, one of then logged a petition to Parliament on this issue, which was later endorsed by more than 3 million people [6]. The government initially declined this petition saying, "*The European Union Referendum Act received Royal Assent in December 2015, receiving overwhelming support from Parliament. The Act did not set a threshold for the result or for minimum turnout.*" However, the government reconsidered this decision and convened a formal discussion of the said petition in early September 2016.

A referendum is a vote used to elucidate the opinion of a country in regards of a particular issue, but it does not presuppose that a difference of one individual in the split should qualify as a majority. The terms of reference for Brexit did not clarify what split would be of *material and practical significance.* The question here is *where should the margin be fixed to claim a majority for either side?* It is relevant to mention that many countries require their referendums to result in larger differences, usually 40-60, for a split to be recognized as of material significance. As we will see in this paper, this apparent "rule of thumb" has a sound scientific basis.

The second issue has to do with the actual calculation of the percentage of vote for Leave and Remain. Quoting 51.9% as the proportion of the population voting for leave, claims are made that the majority of the UK had chosen to leave the EU. On the basis of the 51.9%, statements like "the majority of the UK chose to leave the EU" or "*The British people have voted to leave the European Union*"; "*The will of the British people is…*"; or "*The British people have made a very clear decision to take a different path…*" have pervaded the political discourse. On the one hand, those statements are untrue or, at best, unproven. If there had been a 100% turnout, then we would have complete information about the opinion of the UK population in the Brexit referendum, but this is not the case since the turnout rate was only about 72% which means that only 37% of the eligible British electorate chose Leave (since 51.9% of 72% is 37%. Note that for a 72% turnout a minimum split required to make those claims is 69%-31%). On the other hand, since any statement about the whole population made on the basis of a subset of it comes with an element of uncertainty, we should be compelled to perform the best possible analysis to deduce the views of the whole population from the subset. The uncertainty element should be part of the analysis.

The percentage 51.9% for leave (or 48.1% for remain) was calculated as a simple aggregated average, dividing the total numbers of those voting for leave in the overall population (the UK) by the number of valid votes in the overall population (33,551,983). This method is quick to carry out but, unfortunately, it does bias the results if the population at issue violates the assumption of homogeneity. This assumption, which forms the basis of the method used, did not hold in the population of the Brexit referendum, the United Kingdom. Moreover, this simple aggregated average does not even take into account the within-region variability.

It is not uncommon to fail to foresee heterogeneity in advance of a population study, especially given the size and complexity of a population like the UK which is formed of five different countries and the countries are divided into regions with a diversity of socio-demographics between them. However, given that we did not have a 100% turnout, and given the close result, one should carry out the appropriate analysis to deduce the views of the whole population. In particular, such analysis has to take into account the heterogeneity between the different regions and, most definitely, between countries. Ignoring heterogeneity when it is present biases the results [6].

The aim of this paper is two-fold. Treating the Brexit voters as a sample of responders from the overall UK population,

- First, we assess whether the difference between a 52-48 split (obtained as an overall arithmetic mean in the UK Brexit referendum) and the 50-50 split (representing perfectly even split) is of any *material and practical significance* in a population study as to justify claim of majority for either side.
- Second, we provide valid calculations for the proportion of "leavers" (and, equivalently, of "remainers") in the UK, based on the results obtained in the 2016 Brexit referendum, assessing the *material and practical significance* of the split obtained. In our calculations we take into account, not only the size of the region but also the within-region variability and any between-region heterogeneity that may be present in the population.

We do not get involved with political issues (e.g. choice of the eligible population, manner in which the two campaigns were conducted, benefits or risks brought by any of the sides, etc.). We only concern ourselves with the appropriateness of the calculations performed to decide if there was a majority for Leave, or for Remain; to see if the results are conclusively implying a majority (in either direction) or a tie.

We think there is a bias in the calculation of the Brexit referendum and we want to point out where it lies. Without delving into intricate mathematical exposition, we present the rationale of the methods using a common sense approach, but nevertheless provide references in respect of the mathematical methods used.

**Data sources**

Table 1 exhibits the counts related to the Brexit referendum per each of the five regions that constitute the United Kingdom, according to official sources. The five regions ("countries") are essentially the four countries (England, NI, Scotland and Wales) and Gibraltar. Our sources are the official websites of the Electoral Commission, the BBC and the Daily Telegraph [1, 2, 3, 4]. In the table, column 2, the electorate, shows the number of people eligible to vote in each region. Column 3, the turnout, is the number of people that actually presented to vote. The turnout rate in column 4, is the number of people presenting to vote as a proportion of the electorate. Column 5 shows the number of votes rejected on the basis of scrabbled or double voting. The turnout minus the rejects is the number of valid votes, shown in column 6. Columns 7 and 8 show the number of votes for Remain and Leave respectively. Column 9 shows the proportion of votes for Leave. Given that the vote had a binary outcome, referring to the proportion for leave is sufficient ---the corresponding statement applies for proportion for remain given that proportion for remain is equal to 1 minus the proportion for leave.

**Table 1: Summary of referendum results, by "country" and in the overall UK population**

| Region [1] | Eligible Electorate [2] | Turnout [3] | Turnout rate [4] | Rejects [5] | Valid [6] | Remainers [7] | Leavers [8] | P(leavers) [9] |
|---|---|---|---|---|---|---|---|---|
| England | 38,957,543 | 28,457,414 | 0.73 | 22,157 | 28,435,257 | 13,247,674 | 15,187,583 | 53.4% |
| Gibraltar | 24,119.00 | 20,172.00 | 0.836 | 27 | 20,145 | 19,322 | 823 | 4.1% |
| N Ireland | 1,260,955 | 790,523 | 0.627 | 374 | 790,149 | 440,707 | 349,442 | 44.2% |
| Scotland | 3,987,112 | 2,681,179 | 0.67 | 1,666 | 2,679,513 | 1,661,191 | 1,018,322 | 38.0% |
| Wales | 2,270,272 | 1,628,054 | 0.717 | 1,135 | 1,626,919 | 772,347 | 854,572 | 52.5% |
| **UK** | **46,500,001** | **33,577,342** | **72.20%** | **25,359** | **33,551,983** | **16,141,241** | **17,410,742** | **51.9%** |

All the cited sources were found consistent with each other and all were used in our calculations of results and/or validation of the dataset formats we use. The data is presented in the source tabulated or summarized in different ways, either by countries, regions or areas. The less aggregated form was by area, in a dataset providing complete information for each of 382 areas of the United Kingdom. Each area was identified by name, together with which of the 13 regions of the UK they belong to. The 13 regions are: North East (NE), North West (NW), Yorkshire & The Humber (Yorksh), East Midlands (E Midlands), West Midlands (W Midlands), East, London, South East (SE), South West (SW), Gibraltar, Northern Ireland (NIreland), Scotland and Wales. (We use Midlands to refer to both, West and East Midlands.) The information in this data consisted of the total number of eligible, expected and verified electorate, the number of votes casted and the turnout percentage (already calculated in the dataset), the total of number of valid and rejected votes and the numbers favouring "Leave" and favouring "Remain". In addition, the reasons for the 25,359 rejected votes were provided as "no official" 232, "dual answer" 9,084, "scribbled" 836 and "unmarked" 15,207.

**Analysis Plan**

**Method used to assess if a 48%-52% split in the Brexit referendum imply a majority for leaving the EU.**

In this part, we assess whether a 52-48 split, as the split quoted by the government in the Brexit referendum, is of any material significance in substantiating the claim for a majority of either side. To answer the question of whether the difference observed is large enough to mean a majority in either direction, the difference should be standardized and translated into the so called "effect size", which allows the magnitude of the difference to be assessed, regardless of the units in which the responses were measured in the first place and in a way so that its true magnitude, in relation to the variation across the population, can be assessed [7, 8]. Mathematically, the effect size is the difference of interest divided by the residual standard deviation; the greater the standard deviation the smaller the effect size.

For normally distributed data, the effect size can be expressed in Cohen's *d* scale, which provides a metric in which the difference can be judged as very small, small, medium size, large (as well as negligible or very large) [9]. The scale benchmarks are: a range of 0-0.20 signals a small effect, a range of 0.20-0.50 signals a medium effect and a range of 0.50-1.30 signals a large effect. An effect size of 1.30 and above signals a very large effect, less than 0.10 signals a very small effect and less than 0.05 enters the realm of negligible effects. The log-odds is approximately normally distributed and Chin suggests an efficient method to translate it into the Cohen's d effect size by dividing it by 1.81 [10].

As the responses in the referendum are binary (Leave or Remain) and therefore not normally distributed, the proportion of leavers, say, is transformed into the odds (of leave in relation to remain) and this in turn is transformed into the logarithm of the odds (log-odds). The reason for this is that the log-odds is close to normally distributed, allowing the (standardized) effect size to be assessed using Cohen's *d* scale, which provides benchmarks values to decide whether the difference is negligible, small, medium, large or very large, as explained previously.

Effect size is a concept that has not made a way into the day-to-day toolkit of most political scientists. Likewise for the conceptual difference between *material and practical significance* vs *statistical significance*. In contrast, there is consensus among behavioural, social and biomedical scientists that effect sizes should always be reported and that they are relevant even in cases when statistical hypothesis tests have to be abandoned altogether [7,8]. When the number of observations that intervene in the calculation of a parameter is very large, and not derived on the basis of a power requirement, statistical significance is irrelevant. In contrast, the concept of *material and practical significance* cannot be disregarded.

**Method used to provide a valid calculation of the proportion of "Leavers" in the Brexit referendum**

For this part, we calculate the proportion of leavers in the Brexit referendum, for England alone (divided into its 9 regions) and for the UK. For the latter, the UK is taken as, both, divided into its 13 regions and also divided into its five "countries" (strictly its 4 countries and Gibraltar). For each case, we translate our results (the log-odds) into an effect size, to allow a proper assessment of its magnitude and material significance. The mathematics of our approach is *meta-analytic*, to allow a *valid* aggregation of the results over the different regions, controlling for the sample size in the regions, the variability within the regions and, if present, the variability between regions. (The between-regions variance is usually denoted by *Tau-square*). We refer to the between-region variability as *heterogeneity*.

The naïve aggregate, averaging the head counts over the whole population, takes into account neither the within-region nor the between-region variability. The meta-analytic algorithms provide *valid* aggregated (or pooled), overall, effect sizes (the difference between the overall proportion of leavers and the null value of 50%). We perform our calculations using three different approaches:

- A fixed-effects model, ignoring between-region heterogeneity but giving more weight to those regions with larger sample size and with smaller variability (i.e. more precision). For this purpose the algorithm weights the parameter in each region by the inverse of its variance. For this reason it is known as the *inverse variance fixed effects model* (IV-FE) [8,11]
- A fixed-effects model, still giving more weight to those regions with larger sample size and less variability (using inverse variance weights), but also controlling for the between-region heterogeneity. This (new) algorithm is known as the inverse variance heterogeneity model (IV-Het) [12].
- A random effects (RE) model [13], that aims to improve on the IV-FE model, controlling for the between-region heterogeneity and for this it inverts the total variance (within-region variance plus between-region variance). In this method, if the between-region variability is substantial, it will dominate the denominators of the weights, causing the weights to migrate towards being uniform over the regions. For this reason, given the small size and the status of Gibraltar (a colony, not a country), we present this model with Gibraltar included and excluded.

If no significant between-region heterogeneity is detected, as a minimum, a *fixed-effects* algorithm should be adopted. If heterogeneity is present, it should be adjusted for. In this case, the fixed-effect "might reflect an effect that does not actually exist in the population" [11], and therefore can bias the results.

The heterogeneity assessment is based on Cochran's Q statistic, whose distribution (for our data) can be approximated by a Chi-square distribution with *k-1* degrees of freedom (d.f.), where *k* is the number of regions (or countries) that intervene in the calculation. The heterogeneity assessment is intrinsic to our approach. If significant between-region heterogeneity is found, the results of the regions could be kept separated or, if an overall summary is sought as we know was the case in the Brexit referendum, a model that controls for heterogeneity is used. (The best calculation should take into account the regional heterogeneity).

In the presence of heterogeneity (between regions), the random effects (RE) model incorporates an assumption that the different regions are estimating different, yet related, effects. The existence of heterogeneity suggests that there may not be a single effect but a distribution of effects; this seems to hold in the case of the population subject of the Brexit referendum.

A population can be defined as a (usually) large set of objects of a similar nature that is targeted with the objective of summarizing or elucidating a particular characteristic or state of nature in that set [14]. The concept of population is too often misunderstood or misrepresented. The reason for this is that usually the "population" is associated with the set of objects when in reality it should be associated with the set of objects, each with a label attached, describing the particular attribute or attributes that we are interesting in. A population can be dynamic in some cases or static in other cases. In the case at hand the population is the set of all eligible voters labelled by their opinion: leave, remain or undecided. This dynamic is given by the overcast of uncertainty with respect to voters' decision making. While some voters have already made up their mind to leave or not to leave the EU, many others are "Floating voters" who do not come to a fixed opinion from the outset. Factors contributing to the uncertainty are many, including confusion caused by conflicting political campaign rhetoric. Floating voters bring a dynamic and to a large extent random element into the political picture. A large proportion of these floater voters will be watching the development of the political campaigns being swayed by day-to-day events. A large proportion will be guided by principle rather than loyalty to a particular course. In the case of the Brexit referendum all of this took place in a context of signs of a strikingly atypical phenomena in politics around the world. In the sense that the population of the UK (of individuals with their "labels" e.g. "voting choice") is dynamic and fast changing, the populations (regions), as captured in the referendum, is a random selection. This uncertainty should be a part of the equation in evaluating the results of the Referendum and making final government decision regarding leaving the EU. A model that adjusts for heterogeneity (e.g. the IV-Het and the RE) would quantify this uncertainty.

Let us denote by $P_i$ the statistical populations (regions). Here $i = 1, \cdots, k$ where k may refer to the number of countries (k=5) or regions (k=13). $P_i$ therefore stands for the set of eligible voters in the *ith* region, denoted by $u_j$. In other words $P_i$ is the set *{u1, u2, $\cdots$ }*. For all practical purposes, we can think of $P_i$ as having infinite number of *u's*. Each *u* has a label indicating whether the individual chooses to leave EU or remain. The random effect model has two sources of randomness; the sampling error and the dynamic changes.

For a random sample of size $n_i$ from the population $P_i$, let $Y_{ij}$ = 1 *or* 0 to indicate "leave" or "remain" where $i = 1, \cdots k$ and $j = 1, \cdots, n_i$. Under the static condition, the expected value would be $EY_{ij} = \vartheta_i$, where $\vartheta_i$ is an unknown constant. Under the dynamic condition, we assume $\vartheta$ is a random variable taking values in the set $\Theta = [0, 1]$. We assume a suitable probability distribution on the interval $[0, 1]$, say a beta distribution. (Later in the analysis, by way of logarithmic transformation to effect sizes, we will use normal approximation). Let $\vartheta_i$ be the observed value from $\Theta$ in region *i* after the referendum. The $\vartheta_i$'s are observed independently in each of the k regions. The random effect model assumes that

$$\vartheta_i = \mu + \delta_i \quad \text{for } i = 1, \cdots, k$$

where $\mu$ is the expected value of the beta distribution, the overall proportion of eligible voters in the UK choosing to leave EU, and $\delta_i$ is the random deviation of the *i*th region from the overall $\mu$. (The variance of this random deviation $\delta_i$ is *Tau-square,* the between-regions variance).

We express the proportion of leavers in each region as log-odds. Our approach requires the knowledge of the variability within each region. The data is grouped by area, for the 380 areas in the UK. To obtain the variability in each of the regions, we ungrouped the dataset to have the binary results at each respondent level, for all the 33,578,037 respondents reported in the dataset. Analyses were conducted using STATA V.14 and EXCEL (using the MAXL program [12]).

The results are displayed graphically as "forest plots", where the parameter (effect size) depicted is the overall pooled log-odds of Leave over Remain. In the forest plots, the final results are displayed: the middle point of the rhombus is at the point estimate of the log-odds and the largest (horizontal) diagonal indicates the 95% confidence interval (CI). The relative sizes of the regions is also displayed; indicated by the size of areas of the squares that are placed in the middle of each CI. The forest plots are followed by tables displaying the main summaries with the pooled log-odds (of leave relative to remain), the odds (obtained by "anti-logging" the log-odds) and the proportion of leavers (obtained as P=odds/(1+odds), and the effect sizes obtained via Chin's approximation. The corresponding interpretations are also provided.

It is important to note that since the response to the referendum question was binary (Leave or Remain), it is equivalent to refer to the outcome in three ways: in terms of a split, say 48-50 (as the results quoted in Brexit), or in terms of the proportion of the proportion of voters that chose the "leave" option ("leavers" for short) or in terms of the voters that chose the "remain" option ("remainers" for short). For simplicity of our exposition, and given that the aim of the referendum was to see if the majority was for the leave option, we will be performing the analysis with the outcome expressed in terms of the proportion of "leavers". They are also referred to as the "Brexiters". The terms "leavers", "remainers" and "Brexiters" were adopted in the public domain, by the media, and accepted and used by the population of voters at large themselves.

# Results

**The ungrouped dataset**

The data was grouped at area level, for a total of 382 areas. As explained previously, it was necessary to ungroup the data, from area-level to respondent-level. The ungrouped data resulted in a total of 33,551,983 observations, the number of valid ballots. The following variables were available: a binary indicator on whether each respondent choose Leave or Remain, geographical area and Region. Several checks were run to confirm the correctness of the ungrouped data.

The unadjusted odds in favour of Leave based on the ungrouped data were 1.076911 (SE= 0.000372), which translates into a proportion of 51.9% (0.519), the proportion of respondents choosing Leave that is obtained when the plain arithmetic average of the whole population is calculated irrespective of region. (This consistency among different data formats shows validity of the dataset we are using). The proportion of leavers was transformed to odds and to the corresponding log-odds. Table 2 shows, for each region, the summaries in terms of proportion of leavers, the odds of choosing leave and the log-odds with the corresponding standard deviations.

**Table 2.**
**Summaries calculated by region and in the overall population**

| Country | Region | Areas | Valid | p_leave | Odds | Log-odds | St. Error (log-odds) | LO 95% CI | HI 95% CI. | SD logodds |
|---|---|---|---|---|---|---|---|---|---|---|
| England | NE | 12 | 1,340,698 | 0.58 | 1.383 | 0.324 | 0.00175 | 0.321 | 0.328 | 2.3158 |
| | NW | 39 | 3,665,945 | 0.537 | 1.158 | 0.146 | 0.00100 | 0.144 | 0.148 | 1.9147 |
| | Yorksh | 21 | 2,739,235 | 0.577 | 1.365 | 0.311 | 0.00100 | 0.309 | 0.313 | 1.6551 |
| | E Midlands | 40 | 2,508,515 | 0.588 | 1.428 | 0.356 | 0.00125 | 0.354 | 0.359 | 1.5838 |
| | W Midlands | 30 | 2,962,862 | 0.593 | 1.454 | 0.375 | 0.00125 | 0.372 | 0.377 | 1.7213 |
| | East | 47 | 3,328,983 | 0.565 | 1.298 | 0.261 | 0.00100 | 0.259 | 0.263 | 1.8246 |
| | London | 33 | 3,776,751 | 0.401 | 0.669 | -0.403 | 0.00100 | -0.405 | -0.401 | 1.9434 |
| | SE | 67 | 4,959,683 | 0.518 | 1.074 | 0.071 | 0.00100 | 0.069 | 0.073 | 2.2270 |
| | SW | 37 | 3,152,585 | 0.529 | 1.125 | 0.118 | 0.00125 | 0.115 | 0.12 | 1.7756 |
| England | | 326 | 28,435,257 | 0.534 | 1.146 | 0.156 | 0.00037 | 0.135 | 0.137 | 1.9875 |
| Gibraltar | Gibraltar | 1 | 20,145 | 0.041 | 0.043 | -3.156 | 0.035 | -3.226 | -3.086 | 5.1096 |
| Northern Ireland | N Ireland | 1 | 790,149 | 0.442 | 0.793 | -0.232 | 0.002 | -0.236 | -0.228 | 1.7778 |
| Scotland | Scotland | 32 | 2,679,513 | 0.38 | 0.613 | -0.489 | 0.00125 | -0.492 | -0.487 | 1.6369 |
| Wales | Wales | 22 | 1,626,919 | 0.525 | 1.106 | 0.101 | 0.0015 | 0.098 | 0.104 | 2.5510 |
| UK | | 382 | 33,551,983 | 0.519 | 1.079 | 0.076 | 0.00035 | 0.075 | 0.076 | 1.7377 |

**Would a 48%-52% split quoted by the Brexit referendum imply a majority for either side?**

The difference between a 52%-48% split and a 50%-50% split is summarized expressing it as a difference of 2% between the proportion of leavers (or, equivalently, the proportion of remainers) and 50%. (We can refer to 50% as the null value since it represents the perfectly even split). Therefore the issue is whether a 2% difference (52% vs 50%) is of any material significance. As explained previously, the difference is transformed into the log-odds, the logarithm of the odds of leave (in relation to remain), which in turn is converted to an effect size in Cohen's *d* scale, following Chin's approximation [10], as explained previously.

The split 48.1%-51.9% quoted by the initial calculation in the Brexit referendum (the proportions of respondents choosing Remain being 48.1% and Leave 51.9%) corresponds to an odds of choosing leave (in relation to remain) of 1.079. This mean that the odds of a voter choosing to leave were about 8% greater than choosing to remain. The log-odds is 0.076 and the effect size expressed in the standardized Cohen's d metric is therefore 0.042.

| P (leavers) | Odds of choosing leave | Odds Ratio (*OR*) | log-OR | Effect size (*d*) |
|---|---|---|---|---|
| 0.50 | 1.000 | | | |
| 0.519 | 1.079 | 1.079 | 0.076 | 0.04 |

Contrasting this value against Cohen's *d* scale, this value is in the lowest quarter of the "small" range which constitutes evidence that the material significance of the difference between a 48-52 split and a 50-50 split is, not only small but it falls in the range of negligible to very small. The difference between a 50-50 (perfectly even) split and the 48.1%-51.9% split, quoted by the adjudicators of the Brexit referendum result, is materially, practically negligible.

Using the corresponding inverse transformations, the proportion of leavers (or remainers) required to achieve an effect size is shown in Table 3, which also depicts the effect size benchmarks. We observe that a split of 60%-40% should be the minimum required to indicate a non-small effect size; one that justifies a claim of a majority for either side; a split that is of material significance; a split that would be significantly different from a 50%-50% split in practical or material terms. A split of 55%-45% would indicate a small effect size in practical terms. The observed split 52%-48% translates into an essentially negligible difference in Cohen's classification. The minimum split that is acceptable to declare as of any substantial or material significance should be at least a 60%-40% split. Even a split of 55%-45% would be considered a small difference that could be questioned if used to declare a majority for either side.

Therefore the split 52-48 does not support the claim of a majority for Leave. This also explains why the polls taken a couple of weeks before the referendum showed similar splits but in favour of Remain; such split is spurious and, for this reason, should not form the basis to claim majority for either side.

**Table 3**
**Proportions to achieve a given effect size level**

| Level of effect | Effect size Cohen's *d* | log-OR | Odds Ratio OR | Proportion of leavers |
|---|---|---|---|---|
| Negligible to Very small | <0.05 | 0.0905 | 1.0947 | 0.523 |
| Small-low | 0.10 | 0.1810 | 1.1984 | 0.55 |
| Small | 0.15 | 0.2715 | 1.3119 | 0.567 |
| Small-Medium | 0.20 | 0.3620 | 1.4362 | 0.590 |
| Medium-low | 0.21 | 0.3801 | 1.4624 | 0.594 |
| Medium | 0.30 | 0.5430 | 1.7212 | 0.633 |
| Medium | 0.35 | 0.6335 | 1.8842 | 0.653 |
| Medium-large | 0.50 | 0.9050 | 2.4719 | 0.712 |
| Large | 0.80 | 1.4480 | 4.2546 | 0.810 |
| Large | 1.00 | 1.8100 | 6.1104 | 0.859 |
| Very large | >1.30 | 2.3530 | 10.5171 | 0.913 |

**Calculation of the proportion of Leavers for _England_**

A very highly significant heterogeneity was detected between the 9 regions of England: NE, NW, York & The Humber, East Midlands, West Midlands, East, London, SE and SW. The Cochran measure was Q= 383,792.

The main source of this heterogeneity was London: When London is excluded the Cochran value drastically reduces to Q=79,479. On further inspection, we identified three distinct levels of heterogeneity for England and the 9 English regions were re-grouped into three groups that were relatively homogeneous within (as indicated by the drastic reductions in the Cochran's value), in the following manner:

- The first group was formed by the 5 regions: NE, East, Yorkshire, West Midlands and East Midlands. Their effect sizes were positive and in the upper quarter of the small range of the Cohen's scale, between d=0.14 and d=0.20; their Cochran's value was Q=6,062. The pooled log-odds for this group was 0.32 and the 95% CI was (0.27, 0.36). This corresponds to a Cohen's effect size d=0.16, falling in the upper quarter of the small range in the Cohen's scale: a small size advantage for Leave.
- The second group was formed by the 3 English regions: NW, SW and SE. Their effect sizes were very small for Leave, between d=0.06 and d=0.08; their Cochran's value was Q=2,743. The pooled log-odds for this group was 0.11 and the 95% CI was (0.06, 0.16). This corresponds to a Cohen's effect size d=0.055. This falls between the lowest and second lowest quarters of the small range, indicating a very small advantage for Leave.
- The third group formed by London alone; its Cohen's effect size was d=-0.20. The negative indicates an advantage for Remain; a medium effect size for Remain.

To adjust for the heterogeneity in England we use meta-analytic algorithms, with England divided into these 3 (homogeneous within) regions. Following this, for comparison and completeness of argument, the aggregation of the 9 regions, ungrouped, is also presented. The parameter calculated in all models is the overall log-odds of respondents choosing Leave (over Remain). The results are summarized in Table 4.

**England results with its regions re-grouped into the 3 groups (homogeneous within):**

The forest plots are shown in Figures 1a to 1c. Summaries are displayed in the upper portion of Table 4.

- The fixed effects models weights are: London 13.7%; Yorksh-Midlands-E-NE 44.6% and NW-SE-SW 41.7%. The IV algorithm yields a pooled log-odds of 0.134 with a 95% CI (0.133, 0.135). The positive sign of the point estimate suggests an advantage for Leave. The CI lies completely on the positive axis indicates statistical significance but the effect size falls on the very small portion of the Cohen's scale (as _d_=0.07), indicating that the observed advantage (for Leave) in England is of no material or practical significance. The forest plot is presented in Figure 1a.
- The weights given by the IV-Het algorithm: London 13.7%; Yorksh-Midlands-E-NE 44.6% and NW-SE-SW 41.7%. The IV-Het algorithm yields a pooled log-odds of 0.134 with a 95% CI (-0.24, 0.50). The positive sign of the point estimate suggests an advantage for Leave. The CI contains zero (as the limits have different signs) indicating that there is not statistical significance. The effect size falls on the very small portion of the Cohen's scale (as _d_=0.07), indicating that the observed advantage (for Leave) in England is of no material or practical significance. The forest plot is presented in Figure 1b.
- The random effects (RE) algorithm yields a pooled log-odds of 0.01 with a 95% CI (-0.33, 0.35). The positive sign of the point estimate suggests an advantage for Leave. The CI contains zero (as the limits have different signs) indicating no statistical significance. The effect size falls on the negligible portion of the Cohen's scale (as _d_=0.005), indicating that the observed advantage (for Leave) in England is of no material or practical significance. The forest plot is presented in Figure 1c.

**Figure 1a**
**Forest Plot for the 3 England regions (homogeneous within)**
**Fixed effects (FE), not controlling for heterogeneity [11]**

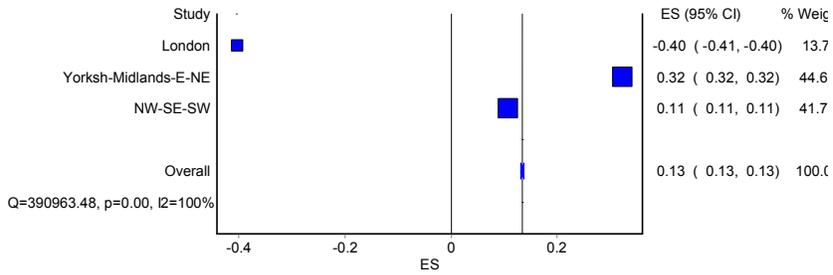

**Figure 1b**
**Forest Plot for the 3 England regions (homogeneous within)**
**Fixed effects controlling for heterogeneity: IV-Het algorithm [11]**

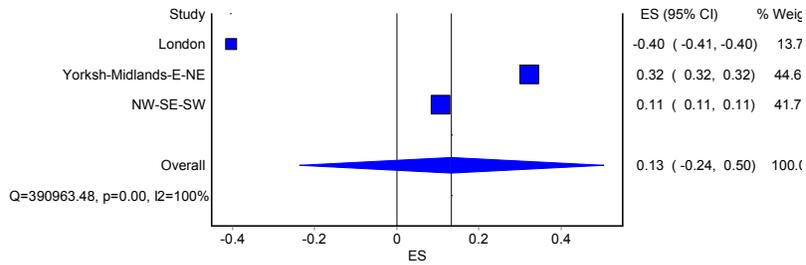

**Figure 1c**
**Forest Plot for the 3 England regions (homogeneous within)**
**Random effects controlling for heterogeneity: RE algorithm [13]**

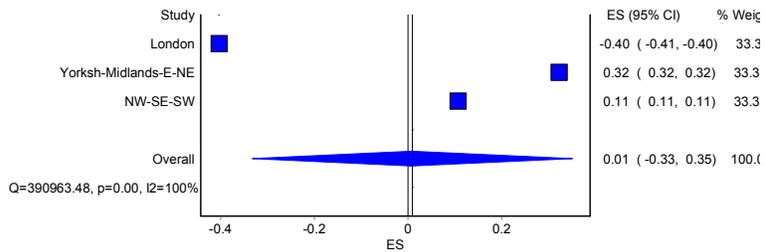

**Table 4**
**Calculations of the actual proportion of leavers in England
controlling for within- and between-region variability**

| England | Log-odds (95% c.i.) | Odds of Leave[1] | Proportion of Leavers[2] | Effect size[3] | Magnitude of difference (in favour of) |
|---|---|---|---|---|---|
| **England divided in 3 regions (homogeneous within)** | | | | | |
| Fixed Effects-Inverse Variance Controls for within-region variability but not for heterogeneity | 0.134 (0.133, 0.135) | 1.14 | 0.53 | 0.067 | Small for LEAVE (*) |
| Fixed effects IV-Het Controls for within-region variability and Heterogeneity | 0.134 (-0.24, 0.50) | 1.14 | 0.53 | 0.067 | Small for LEAVE (NS) |
| RE Random Effects Controls for within-region variability and Heterogeneity | 0.01 (-0.33, 0.35) | 1.01 | 0.502 | 0.005 | Negligible for LEAVE (NS) |
| **England divided in 9 regions** | | | | | |
| Fixed Effects-Inverse Variance Controls for within-region variability but not for heterogeneity | 0.1375 (0.137, 0.138) | 1.147 | 0.534 | 0.076 | Small for LEAVE |
| Fixed effects IV-Het Controls for within-region variability and Heterogeneity | 0.14 (-0.03, 0.31) | 1.150 | 0.54 | 0.08 | Small for LEAVE (NS) |
| RE Random Effects Controls for within-region variability and Heterogeneity | 0.17 (0.01, 0.34) | 1.185 | 0.54 | 0.09 | Small for LEAVE (*) |

[1]Odds of leave=exp(log-odds); [2]proportion of leavers=(odds of leave/(1+odds of leave); [3]effect size=log-odds/1.81 (Chin, 2000 [10])

**England results with its 9 regions (ungrouped)**

The forest plots are shown in Figures 2a to 2c. Summaries are displayed in the lower portion of Table 4.

For comparison, the results for England alone with the 9 regions ungrouped, are also summarized in Table 4 (lower portion). We observe that controlling for both, the within and between region variabilities, the proportion of leavers in England is 54%. The log-odds is positive (in both models), indicating an advantage for Leave. This advantage represents a very small difference: the effect size is about 0.08 to 0.10 in Cohen's scale, hence it is of no *material or practical* significance. The *statistical significance* can be judged from the forest plots according to whether zero is contained in the CI: in the fixed effects IV-Het algorithm that adjusts for heterogeneity (Figure 2a), the difference does not reach statistical significance. In the random effects algorithm [13], which also controls for heterogeneity (Figure 2b), the difference is statistically significant. However, although the 95% CI falls entirely within the positive axis (indicating statistical significance), it is very wide, indicating the low precision of the estimate and, moreover, the CI is not well removed from zero.

**Figure 2a**
**Forest Plot for the 9 regions of England**
**Fixed effects model controlling for between-region heterogeneity**
**(IV-Het algorithm[12])**

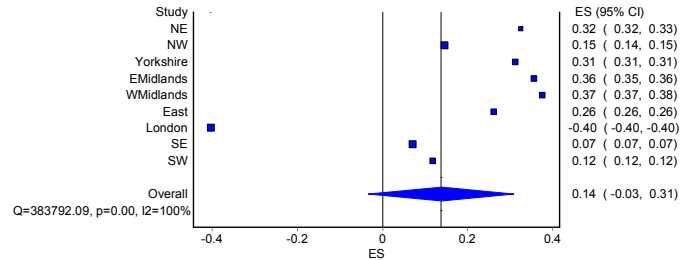

NB. The pooled log-odds (ES) here is positive suggesting, for England, an advantage for Leave.

**Figure 2b**
**Forest Plot for the 9 regions of England**
**Random effects (RE) model controlling for between-region heterogeneity [13]**

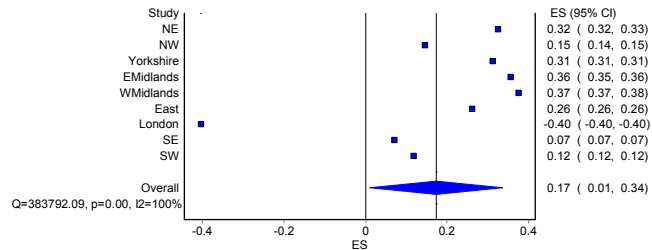

NB. The pooled log-odds (ES) here is positive suggesting, for England, an advantage for Leave.

**Calculation of the proportion of Leavers for the _United Kingdom_**

There was a highly significant heterogeneity between the 13 geographical/political regions of the UK (Wales, Scotland, Northern Ireland, Gibraltar and the 9 regions of England). For the 13 regions, the Cochran value is Q= 200,000 and the I-squared indicated that essentially all variation in effect sizes was attributable to between country heterogeneity. Fitting England as a country (instead of divided in its 9 regions) reduces the Cochran value to Q= 183,178 --with Gibraltar included, and to Q=175,416 --with Gibraltar excluded. These values still represent very highly significant heterogeneity and, in both cases, the I-sq statistic continues indicating that most of the variability is caused by the substantial variability between the countries (the between-region variance was estimated as 0.08). Heterogeneity should not be ignored.

a. **Taking the UK divided in 6 regions (with the 9 England regions grouped into the 3 regions relatively homogeneous within.**

We fit the models to the UK, divided into six regions: the 3 English (relatively homogeneous within) regions plus Northern Ireland, Scotland, Wales. Figures 3a and 3b show the forest plots for, respectively, the fixed effects model that adjust for heterogeneity, the IVhet algorithm [12] and the random effects model [13]. Gibraltar is excluded from these models. With the six regions, Cochran's statistic was Q=633,790. Table 5 summarizes the results.

- The point estimate of the log-odds is 0.08 under the IV-het algorithm. The positive sign in the log-odds suggests an advantage for Leave. This corresponds to an odds (for leave relative to remain) of 1.08 and a proportion of leavers of 51.9%. In the Cohen's scale this corresponds to an effect of size *d*=-0.038, a *very small* advantage for Leave and not statistically significant (as zero is contained in the 95% CI). Under the RE algorithm the log-odds is -0.10 which in Cohen's scale is a small advantage for Remain (as *d*=-0.05) and not statistically significant (as zero is contained in the 95% CI).
- The previous models do not include Gibraltar. When Gibraltar is included (in the London group to optimize homogeneity within), Cochran's value changes slightly is Q=627,894. The pooled log-odds for this group was -0.10 and the 95% CI was (-3.62, 0.162), with the negative sign of the log-odds suggesting the advantage is for Remain. The corresponding Cohen's effect size is d=-0.05, a small advantage for Remain and not statistically significant (as zero belongs to the 95% CI).

**Figure 3a**
**Fixed Effects heterogeneity adjusted IVhet algorithm [12]**

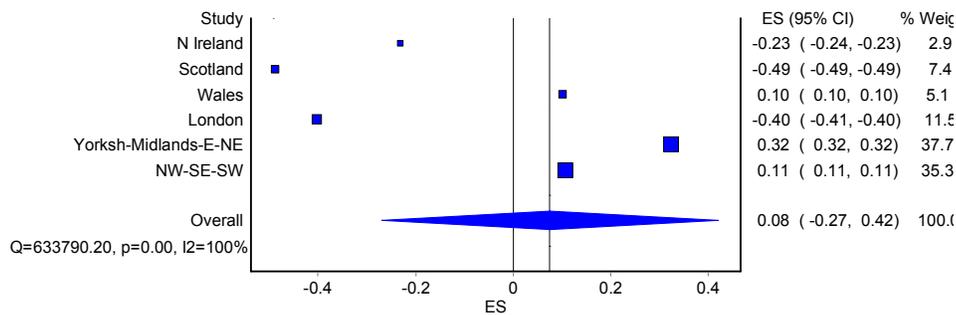

**Figure3b**
**Random Effects - heterogeneity adjusted [13]**

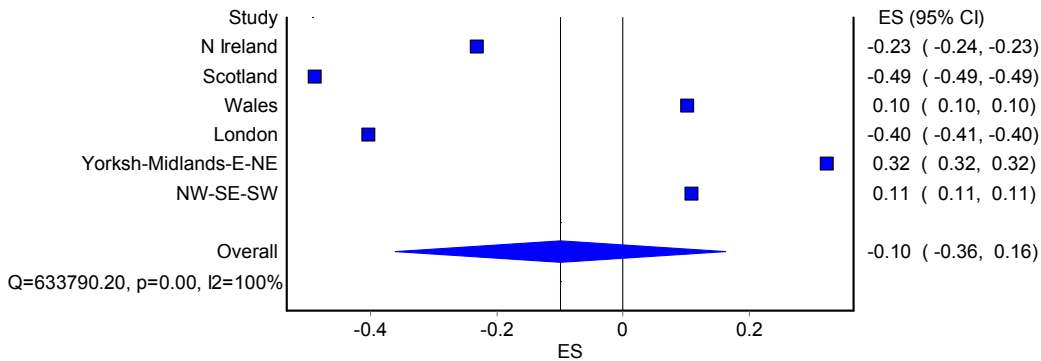

**Table 5**
**Calculations of the actual proportion of leavers in the UK (6 regions) by different methods.**

| England | Log-odds (95% c.i.) | Odds of Leave[1] | Proportion of Leavers[2] | Effect size[3] $d$ | Difference in Cohen's scale (for which campaign) |
|---|---|---|---|---|---|
| FE model (INVHet) Controls for within-country variability. Adjusts for between-country heterogeneity | 0.076 (-0.27, 0.421) | 1.08 | 0.519 | 0.038 | Very small for Leave (NS) |
| RE (Gibraltar excluded) Controls for within-country variability and Heterogeneity inverting the total variance | -0.10 (-0.36, 0.16) | 0.905 | 0.475 | -0.05 | Small for Remain (NS) |
| RE model (Gibraltar included) Controls for within-country variability and Heterogeneity inverting the total variance | -0.41 (-4.22, 3.41) | 0.66 | 0.40 | -0.20 | Medium size for REMAIN (NS) |

b. **Taking the UK divided in its "countries"**

The point estimate as well as the 95% CI for the overall (pooled) log-odds are provided. The log-odds is expressed in Cohen's scale to allow an assessment of the size of the magnitude of the difference. The forest plots are presented in Figures 4a to 4c. The summaries are presented in Table 6. We exhibit both models, including and excluding Gibraltar for completeness of argument (especially in the case of the RE algorithm, see Figure 4c).

- The fixed-effects inverse variance algorithm that adjusts for heterogeneity IV-Het [12] yields an overall log-odds (ES) of 0.04 (95% CI -0.47 to 0.56). (The same results are obtained when Gibraltar is excluded. Note that a substantial 75.6% of the weight is apportioned to England and essentially no weight is apportioned to Gibraltar.). The point estimate of the log-odds is positive, suggesting an advantage for Leave. However, in the UK, this apparent advantage for "Leave" is not substantiated by this model, as implied by the *negligible to very small* effect size in Cohen's scale, *d*=0.02. This advantage is not only of no material or practical significance: since zero is not contained in the 95% CI, it is also not statistically significant.
- The random-effects (RE) algorithm when Gibraltar is excluded, see Figure 4b, yields a log-odds of -0.12 (95% CI -0.45 to 0.21). The negative sign of the point estimate now indicates an advantage for Remain. Since the effect size in Cohen's scale is -0.06, it represents a *small* effect. The CI contains zero (see the different signs in the limits) which indicates that this small advantage is also not statistically significant. However, we notice that the CI is well shifted towards the negative axis, providing evidence for Remain.
- The random-effects (RE) algorithm [13] that weights the parameters according to the total variance (within and between countries), see Figure 4c, yields a log-odds of -0.72 (95% CI -1.0 to -0.42). The negative sign in the point estimate suggests an advantage for Remain. This advantage is substantial: using Chin's approximation, the standardized effect size in Cohen's scale is *d*=-0.36, a medium-large effect ---it falls in the upper portion of the range for medium size effects. The negative signs in both limits of the CI, indicate that zero is not contained in the CI and imply that this medium-large advantage observed for Remain, when Gibraltar is included, is also statistically significant.

**Figure 4a**
**Forest Plot for the 5 countries of the UK**
**Fixed effects model controlling for heterogeneity - IVhet algorithm [12]**

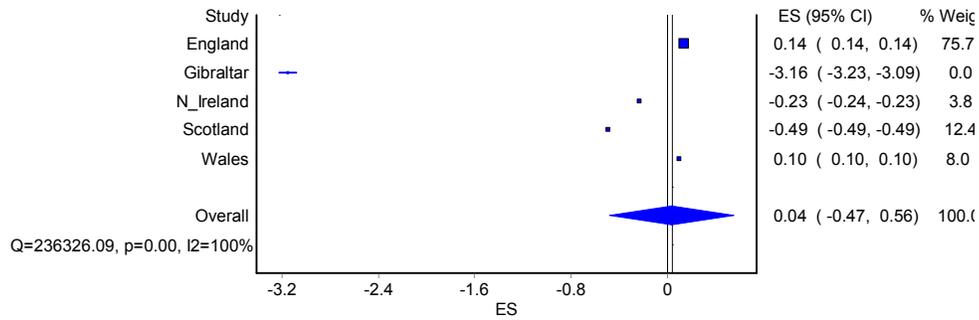

**Figure 4b:**
**Random Effects Forest Plot for the 4 countries of the UK - Controlling for heterogeneity [13]**
**(Gibraltar excluded)**

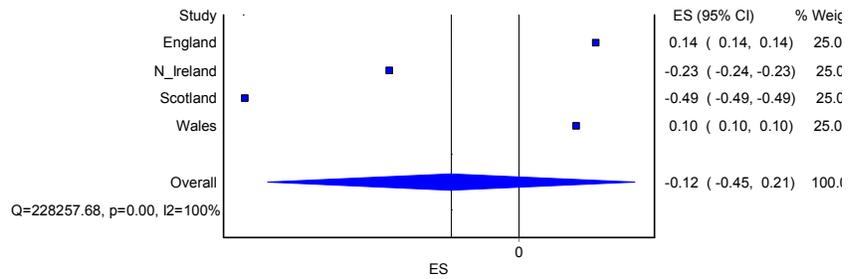

**Figure 4c:**
**Random Effects Forest Plot for the "countries" of the UK - Controlling for heterogeneity [13]**
**(Gibraltar included)**

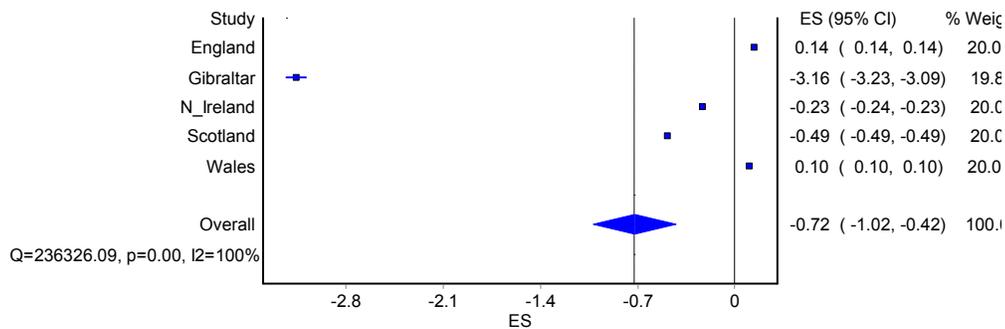

**Table 6**
**Calculations of the actual proportion of leavers in the UK (4 "countries") by different methods.**

| England | Log-odds (95% c.i.) | Odds of Leave[1] | Proportion of Leavers[2] | Effect size[3] $d$ | Difference in Cohen's scale (for which campaign) |
|---|---|---|---|---|---|
| IV-Het Controls for within-region variability and Heterogeneity | 0.04 (-0.47, 0.56) | 0.90 | 0.475 | 0.02 | Negligible to very small for Remain (NS) |
| RE Random Effects (Gibraltar excluded) Controls for Heterogeneity | -0.12 (-0.45, 0.21) | 0.90 | 0.475 | -0.06 | Small For Remain (NS) |
| RE (Gibraltar included) Controls for Heterogeneity | -0.72 (-1.0, -0.42) | 0.487 | 0.327 | -0.36 | Medium-large for REMAIN (*) |

[1]Odds of leave=exp(log-odds of leave); [2]proportion of leavers=(odds of leave/(1+odds of leave); [3]effect size=log-odds/1.81 (Chin, 2000 [10])

### b. Taking the *United Kingdom* divided in 13 regions

The Cochran heterogeneity statistics for the 13 regions was Q=639,057 and the between-region variance was =0.09. The I-sq statistic indicated that most of the heterogeneity was due to the between-region variability. This heterogeneity summary showed to be most sensitive to Scotland and London. When either of these two regions is excluded, separately, the Cochran statistics is substantially reduced, to 418,500 and 403,000 respectively. Due to the relatively small size, excluding Gibraltar showed to have no effect on the heterogeneity (Q is slightly reduced, to 630,797). The meta-analytic models are fitted and the forest plots are shown in Figures 5a to 5d. Table 7 summarizes the results. The interpretation of these results is as follows:

- The fixed-effects inverse variance algorithm that adjusts for heterogeneity, IV-Het [12], see Figure 6a, yields an overall log-odds (ES) of 0.08 (95% CI -0.10 to 0.26). The point estimate of the log-odds is positive, suggesting an advantage for Leave. However, this apparent advantage for "Leave" is not implied by this model since the effect size in Cohen's scale is negligible to very small, d=0.044. We also observe that zero is contained in the confidence interval indicating that this advantage is not statistically significant. Note that a substantial 75.6% of the weight is apportioned to England and essentially no weight is apportioned to Gibraltar and, in fact, the same results are obtained when Gibraltar is excluded.
- The random-effects (RE) model [13], when Gibraltar is excluded (see Figure 6b), yields a log-odds of 0.08 (95% CI -0.009 to 0.24). The positive sign of the point estimate suggests an advantage for Leave with a negligible to very small magnitude: The effect size in Cohen's scale is 0.04. This is an effect size of neither material significance nor statistical significance (as zero is contained in the confidence interval).
- The random-effects (RE) model [13], see Figure 6c, yields a log-odds of -0.17 (95% CI -0.33 to -0.01) when Gibraltar is included. The negative sign in the point estimate indicates an advantage for Remain. The negative signs in both limits of the confidence interval, indicate that that zero is not contained in the confidence interval and therefore the advantage for Remain is statistically significant. However, although statistically significant, this advantage is of only small material or practical significance: the effect size in Cohen's scale is d=-0.09, a small effect ---it falls in the middle of the range for small size effects.

### Figure 5a
### Forest Plot for the 13 regions of the UK
### Fixed effects IV-Het algorithm adjusting for heterogeneity [12]

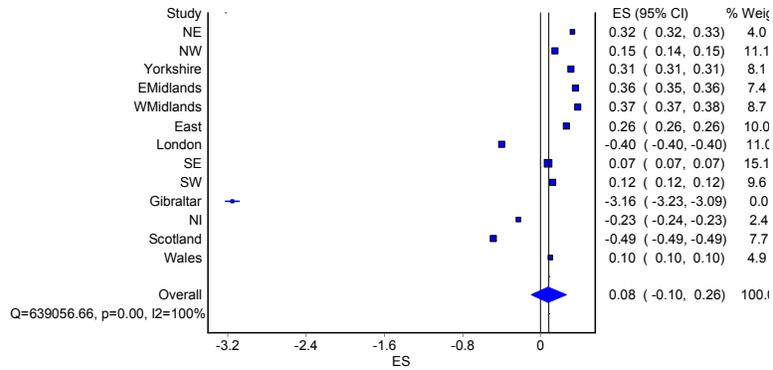

### Figure 5b
### Random Effects Forest Plot for the 13 regions of the UK
### Controlling for between-country heterogeneity [13]
### Gibraltar excluded

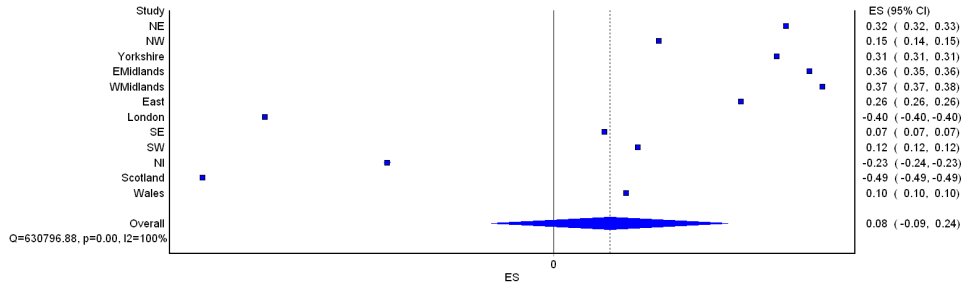

### Figure 5c
### Random Effects Forest Plot for the 13 regions of the UK
### Controlling for between-country heterogeneity [13]
### Gibraltar included

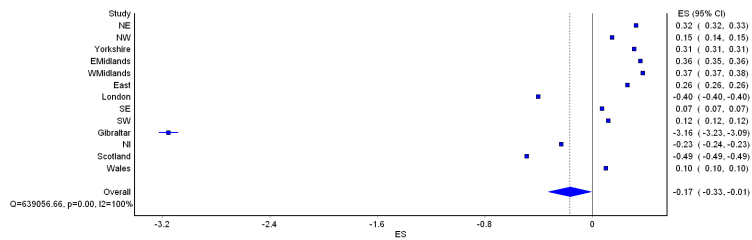

**Table 7**
**Calculations of the actual proportion of leavers in the UK (13 regions)
controlling for both, within- and between-region variability**

| England | Log-odds (95% c.i.) | Odds of Leave[1] | Proportion of Leavers[2] | Effect size[3] | Magnitude of difference (in favour of) |
|---|---|---|---|---|---|
| FE model (INVHet) Controls for within-country variability. Adjusts for between-country heterogeneity | 0.08 (-0.10, 0.26) | 1.08 | 0.519 | 0.04 | Very small For LEAVE (NS) |
| Random Effects (Gibraltar excluded) Controls for within-region variability and Heterogeneity through inverse total variance | 0.08 (-0.09, 0.24) | 1.08 | 0.519 | 0.04 | Very small For LEAVE (NS) |
| Random Effects (Gibraltar included) Controls for within-region variability and Heterogeneity through inverse total variance | -0.17 (-0.33, -0.01) | 0.846 | 0.46 [1] | -0.09 | Small-medium for REMAIN (*) |

[1]Odds of leave=exp(log-odds of leave); [2]proportion of leavers=(odds of leave/(1+odds of leave); [3]effect size=log-odds/1.81 (Chin, 2000 [10])

**General observations about the overall message based on the results obtained by the different algorithms that adjust for heterogeneity.**

The better the adjustment for the highly significant heterogeneity, the further the aggregation moves away from the Leave side, showing advantage for Remain in many instances.

There is a remarkable consistency in the results for England and the UK when England is ungrouped or re-grouped into regions of relative homogeneity within.

All algorithms consistently suggest that, at the most, there is no advantage of substantial material or practical importance for either side.

**Discussion**

Heterogeneity could have been anticipated in advance of the Brexit referendum. However, it is not uncommon to miss the heterogeneity in a population study and, for this reason, the assessment of whether or not heterogeneity is present should constitute the first step in the analysis of the data, before any population parameter is calculated. If significant variability between regions is detected, the mathematics used for the calculation of the proportion of Leave voters (or Remain voters) needs to be slightly more sophisticated that a simple arithmetic average of the aggregated population. When aggregating data in a population that has natural clusters, the random effects model should be the approach of choice. If there is no heterogeneity there will not be any risk in taking this approach as the results of the random effects approach will coincide with those of the fixed effects approach. For all this reasons, social scientists and investigators recommend that the analysis adopted should be one that takes into account heterogeneity [13, 14, 15]. Moreover, in the Brexit referendum, the results obtained under the random effects approach are different from those obtained under the fixed effects approach, which means that there was a large between-regions heterogeneity that, simply, cannot be ignored.

It would be an interesting scientific project to consider the age, income and education distributions of the 380 areas, to try to identify what factor(s) drove the vote in the direction it went. Although this would be an exercise with a valuable sociological research objective and worthy of future research, our purpose here was only to elucidate whether or not the majority of voters favour the leave the EU option and to obtain the correctly aggregated summary, the proportion of leavers or remainers based on the Brexit data.

On the different topic of whether the 52%-48% split represents a meaningful, non-negligible majority for one campaign, our paper demonstrates that the 52%-48% split (corresponding to a difference of 2% relative to the null value at issue 50%; as observed in the referendum) should not have "qualified" as a conclusive result on which to found the claim that the majority of the UK had chosen Brexit. The observed split 48%-52% translates to an essentially negligible difference in Cohen's classification.

The fact that the polls predicted a 52%-48% split with an advantage for Remain a week before the actual referendum and the actual result was 52%-48% with the advantage for Leave reflects the fact that such a split is of negligible importance, a spurious one.

Where should the margin be fixed to claim a majority result in a (binary) referendum? As explained in our paper, Cohen's *d* scale provides a framework to judge whether the proportion of leave voters is negligible, small, medium size, large or very large according to established effect sizes benchmark values.

In this age of data analytics and evidence-based practices, a sound and responsible statistical evaluation of any referendum that is going to influence public policy is of paramount importance. We contend that a referendum is primarily a population study that is going to be used to make claims like "*the UK population has voted to leave the EU*". Such claims are being articulated frequently by political leaders but they are not supported by a rational analysis.

We have analysed the published data of the referendum and, leaving aside those issues associated with planning of this exercise (for example, the choice of which sectors of the population are eligible), our results indicate two fundamental problems. First, the split quoted by the government, albeit questionable, is not a split of any material significant difference with a 50%-50% split and therefore does not support the claim that the majority of the voters chose to leave the EU. The effect size (standardized difference) lies on the lowest quarter of the small range in the effect size scale. The other issue is that crucial regional heterogeneity has been ignored and therefore not factored in the calculations. When this is taken into account, the results are different, as shown above, and they indicate a tie for the two campaigns.

## Conclusion

The 52%-48% split does not represent any substantial, material or practical significant difference in relation to the 50%-50% split (the even split) and should not be used to declare a majority for either side or campaign. The claim that the majority of the respondents of the UK voted in favour of leaving the EU is not substantiated by any of the mathematical calculations: including the simple aggregation adopted on the night of the referendum and any of the methods that control for the within region variability and the between region heterogeneity present in the data of the Brexit referendum.

On using the appropriate methods, we found that the more we adjust for the heterogeneity present in the Brexit referendum, the further the results move away from Leave, albeit still within the range of *small* standardized differences. All differences are consistently found to be of essentially neither of <u>*material or practical*</u> significance nor of <u>*statistical*</u> significance. Therefore, the data does not support majority for either side. It, essentially, signals a tie between the two campaigns.

## Acknowledgements

The authors are grateful to Sir Simon Donaldson, Professor at Imperial College, for reading drafts of the paper and for his very helpful comments and suggestions.

# References

[1] http://www.electoralcommission.org.uk/find-information-by-subject/elections-and-referendums/upcoming-elections-and-referendums/eu-referendum/electorate-and-count-information

[2] http://www.bbc.co.uk/news/uk-politics-36616028

[3] http://www.telegraph.co.uk/news/2016/06/23/leave-or-remain-eu-referendum-results-and-live-maps/

[4] http://www.bbc.co.uk/news/politics/eu_referendum/results

[5] https://petition.parliament.uk/petitions/131215?reveal_response=yes. Last assessed on 28/07/2016

[6] Cumming, G. (2012). Understanding the New Statistics: Effect Sizes, Confidence Intervals, and Meta-Analysis. New York: Taylor & Francis.

[7] Kirk, R. (1996). Practical significance: A concept whose time has come. Educational and Psychological Measurement, 56, 746-759.

[8] Kline, R. B. (2013). Beyond Significance Testing: Statistics Reform in the Behavioral Sciences. 2nd ed. Washington, DC: American Psychological Association.

[9] Jacob Cohen *(1988)*. Statistical Power Analysis for the Behavioral Sciences (second ed.). Lawrence Erlbaum Associates.

[10] Chin S. A simple method for converting an odds ratio to effect size for use in meta-analysis. Statist. Med. 2000; 19:3127-3131. http://www.aliquote.org/pub/odds_meta.pdf

[11] Cochrane Collaboration handbook. Strategies for addressing heterogeneity. http://handbook.cochrane.org/chapter_9/9_5_3_strategies_for_addressing_heterogeneity.htm . Last assessed on 05/08/2016 pm.

[12] Doi SA, Barendregt S, Khan L and GM Williams. (2015) "Advances in the meta-analysis of heterogeneous clinical trials I: The inverse variance heterogeneity model. Contemp Clin Trials 45: 130-138

[13] DerSimonian R, Laird N. Meta-analysis in Clinical Trials. Controlled Clinical Trials 1986; 7:177-188.

[14] http://www.statistics.com/glossary&term_id=812 . Last retrieved on 07/08/2016

[15] Ades AE, Higgins JPT. The Interpretation of Random-Effects Meta-Analysis in Decision Models. Medical Decision Making 2005; 25:646-654.